\documentclass[fleqn,10pt]{wlscirep}
\usepackage[utf8]{inputenc}
\usepackage[T1]{fontenc}
\usepackage{tikz}
\usepackage{multirow}
\usepackage{ctable}
\usepackage{subcaption}
\usepackage{todonotes}
\title{The COUGHVID crowdsourcing dataset: A corpus for the study of large-scale cough analysis algorithms}

\author[1,$\dag$]{Lara Orlandic}
\author[1,$\dag$]{Tomas Teijeiro}
\author[1]{David Atienza}
\affil[1]{Embedded Systems Laboratory (ESL), EPFL, Lausanne, 1015-Switzerland}

\affil[*]{corresponding author: Lara Orlandic (lara.orlandic@epfl.ch)}

\affil[$\dag$]{these authors contributed equally to this work}

\begin{abstract}

Cough audio signal classification has been successfully used to diagnose a variety of respiratory conditions, and there has been significant interest in leveraging Machine Learning (ML) to provide widespread COVID-19 screening. However, there is currently no validated database of cough sounds with which to train such ML models. The COUGHVID dataset provides over 20,000 crowdsourced cough recordings representing a wide range of subject ages, genders, geographic locations, and COVID-19 statuses. First, we filtered the dataset using our open-sourced cough detection algorithm. Second, experienced pulmonologists labeled more than 2,000 recordings to diagnose medical abnormalities present in the coughs, thereby contributing one of the largest expert-labeled cough datasets in existence that can be used for a plethora of cough audio classification tasks. Finally, we ensured that coughs labeled as symptomatic and COVID-19 originate from countries with high infection rates, and that their expert labels are consistent. As a result, the COUGHVID dataset contributes a wealth of cough recordings for training ML models to address the world’s most urgent health crises.

\end{abstract}
\begin{document}

\flushbottom
\maketitle

\thispagestyle{empty}

\iffalse
\noindent Please note: Abbreviations should be introduced at the first mention in the main text – no abbreviations lists or tables should be included. Structure of the main text is provided below.
\fi

\section*{Background \& Summary}

The novel coronavirus disease (COVID-19), declared a pandemic by the World Health Organization on March 11, 2020, has claimed over 950,000 lives worldwide as of September 2020 \cite{covidStats}. Epidemiology experts concur that mass testing is essential for isolating infected individuals, contact tracing, and slowing the progression of the virus \cite{Rosenthal2020,Salathe2020}. While advances in testing have made these tools more widespread in recent months, there remains a need for inexpensive, rapid, and scalable COVID-19 screening technology \cite{MacKay2020}.

One of the most common symptoms of COVID-19 is a dry cough, which is present in approximately 67.7\% of cases \cite{whochina}. Cough sound classification is an emerging field of research that has successfully leveraged signal processing and artificial intelligence (AI) tools to rapidly and unobtrusively diagnose respiratory conditions like pertussis \cite{Pramono2016}, pneumonia and asthma \cite{Amrulloh2015} using nothing more than a smartphone and its built-in microphone. Several research groups have begun developing algorithms for diagnosing COVID-19 from cough sounds \cite{Imran, Brown}. One such initiative, AI4COVID \cite{Imran}, provides a proof-of-concept algorithm but laments the lack of an extensive, labeled dataset that is needed to effectively train Deep Learning (DL) models.

There are several existing COVID-19 cough sound datasets used to train Machine Learning (ML) models. Brown et al.\cite{Brown} have amassed a crowdsourced database of more than 10,000 cough samples from 7,000 unique users, 235 of which claim to have been diagnosed with COVID-19. However, the authors have not automated the data filtering procedure and consequently needed to endure the time-consuming process of manually verifying each recording. Furthermore, this dataset is not yet publicly available and therefore cannot be used by other teams wishing to train their ML and DL models. The Coswara project\cite{Coswara}, on the other hand, has publicly provided manual annotations of the crowdsourced COVID-19 coughs they have received, but as of September 2020, their dataset contains just slightly over 1,3000 samples. An alternative approach to crowdsourcing is the NoCoCoDa\cite{Cohen-McFarlane2020}, a database of cough sounds selected from media interviews of COVID-19 patients. However, this database only includes coughs from 10 unique subjects, which is not enough for AI algorithms to successfully generalize to the global population.

In this work, we present the COUGHVID crowdsourcing dataset, which is an extensive, validated, and publicly-available dataset of cough recordings. With more than 20,000 recordings -- 1,010 claiming to have COVID-19 -- originating from around the world, it is the largest known COVID-19-related cough sound dataset in existence. In addition to publicly providing our cough corpus, we have trained and open-sourced a cough detection ML model to filter non-cough recordings from the database. Furthermore, we have undergone an additional layer of validation whereby 3 expert pulmonologists annotated a fraction of the dataset to determine which crowdsourced samples realistically originate from COVID-19 patients.

In addition to COVID-19 diagnoses, our expert labels and metadata provide a wealth of insights beyond those of existing public cough datasets. These datasets either do not provide labels or contain a small number of samples. For example, the Google Audio Set\cite{Gemmeke} contains 871 cough sounds, but it does not specify the diagnoses or pathologies of the coughs. Conversely, the IIIT-CSSD\cite{Singh2016} labels coughs as wet vs dry and short-term vs long-term ailments, but it only includes 30 unique subjects. The COUGHVID dataset contributes over 2,000 expert-labeled coughs, all of which provide a diagnosis, severity level, and whether or not audible health anomalies are present, such as dyspnea, wheezing, and nasal congestion. Using these expert labels along with subject metadata, our dataset can be used to train models that detect a variety of subjects' information based on their cough sounds. Overall, our dataset contains samples from a wide array of subject ages, genders, COVID-19 statuses, pre-existing respiratory conditions, and geographic locations, which potentially enable AI algorithms to successfully perform generalization. 

Finally, we assert the validity of our data by ensuring that samples labeled as COVID-19 originate from countries where the virus was prevalent at the time of recording, and that the experts exhibit a reasonable degree of agreement on the cough diagnoses. The first step to building robust AI algorithms for the detection of COVID-19 from cough sounds is having a reliable, high-quality dataset, and the COUGHVID dataset effectively meets this pressing global need.

\iffalse
(700 words maximum) An overview of the study design, the assay(s)
performed, and the created data, including any background information
needed to put this study in the context of previous work and the literature.
The section should also briefly outline the broader goals that motivated
the creation of this dataset and the potential reuse value. We also
encourage authors to include a figure that provides a schematic overview
of the study and assay(s) design. The Background \& Summary should not include subheadings. 
This section and the other main body sections of the manuscript should include citations to 
the literature as needed. 
\fi

\section*{Methods}

% The Methods should include detailed text describing any steps or procedures used in producing the data, including full descriptions of the experimental design, data acquisition assays, and any computational processing (e.g. normalization, image feature extraction). See the detailed section in our submission guidelines for advice on writing a transparent and reproducible methods section. Related methods should be grouped under corresponding subheadings where possible, and methods should be described in enough detail to allow other researchers to interpret and repeat, if required, the full study. Specific data outputs should be explicitly referenced via data citation (see Data Records and Citing Data, below).
% 
% Authors should cite previous descriptions of the methods under use, but ideally the method descriptions should be complete enough for others to understand and reproduce the methods and processing steps without referring to associated publications. There is no limit to the length of the Methods section. Subheadings should not be numbered.
% 
% \subsection*{Subsection}
% 
% Example text under a subsection. Bulleted lists may be used where appropriate, e.g.
% 
% \begin{itemize}
% \item First item
% \item Second item
% \end{itemize}
% 
% \subsubsection*{Third-level section}
%  
% Topical subheadings are allowed.

\subsection*{Data Collection}
All of the recordings were collected between April 1st, 2020 and September 10th, 2020 through a Web application deployed on a private server located at the facilities of the École Polytechnique Fédérale de Lausanne (EPFL), Switzerland. The application was designed with a simple workflow and following the principle ``one recording, one click'', according to which if someone simply wants to send a cough recording, they should have to click on no more than one item. 

The main Web interface  has just one ``Record'' button that starts audio recording from the microphone for up to 10 seconds. Once the audio recording is completed, a small questionnaire is shown to get some metadata about the age, gender, and current condition of the user, but even if the questionnaire is not filled, the audio is sent to the server. The variables captured in the questionnaire are described in Table~\ref{tab:metadata}. Also, the user is asked for permission to provide their geolocation information, which is not mandatory. Finally, since coughing is a potentially dangerous activity in the scope of a global pandemic, we provide easy-to-follow safe coughing instructions, such as coughing into the elbow and holding the phone at arm's length, that can be accessed from the main screen.

\subsection*{Database Cleaning}
A common pitfall of crowdsourced data is that it frequently contains samples unrelated to the desired content of the database. In order to allow users of the COUGHVID database to quickly exclude non-cough sounds from their analyses, we developed a classifier to determine the degree of certainty to which a given recording constitutes a cough sound. These output probabilities of the classifier were subsequently included in the metadata of each record under the \texttt{cough\_detected} entry.

To train the classifier, we first hand-selected a set of 121 cough sounds and 94 non-cough sounds including speaking, laughing, silence, and miscellaneous background noises. These recordings were preprocessed by lowpass filtering ($f_{cutoff} = 6$ kHz) and downsampling to 12 kHz. Next, 68 audio features commonly used for cough classification were extracted from each recording. 40 of the features are those described by Pramono et al. \cite{Pramono2016}, and the implementation of our 19 Energy Envelope Peak Detection features is detailed by Chatrzarrin et al. \cite{Chatrzarrin2011}. We also included a signal length feature, as well as the power spectral density (PSD) of the signal in 8 hand-selected frequency bands. These bands were chosen by analyzing the PSDs of cough vs non-cough signals and selecting the frequency ranges with the highest variation between the two classes. Finally, we trained an eXtreme Gradient Boosting (XGB)\cite{XGB} classifier using 78\% of the available data. The hyperparameters of the XGB model were tuned using Tree-structured Parzen Estimators (TPE) \cite{NIPS2011_4443} with a precision objective using the training data for 10-fold, 20\%-test shuffle-split cross-validation. 

\begin{figure}[ht]
  \centering
  % include first image
  \includegraphics[width=0.8\linewidth]{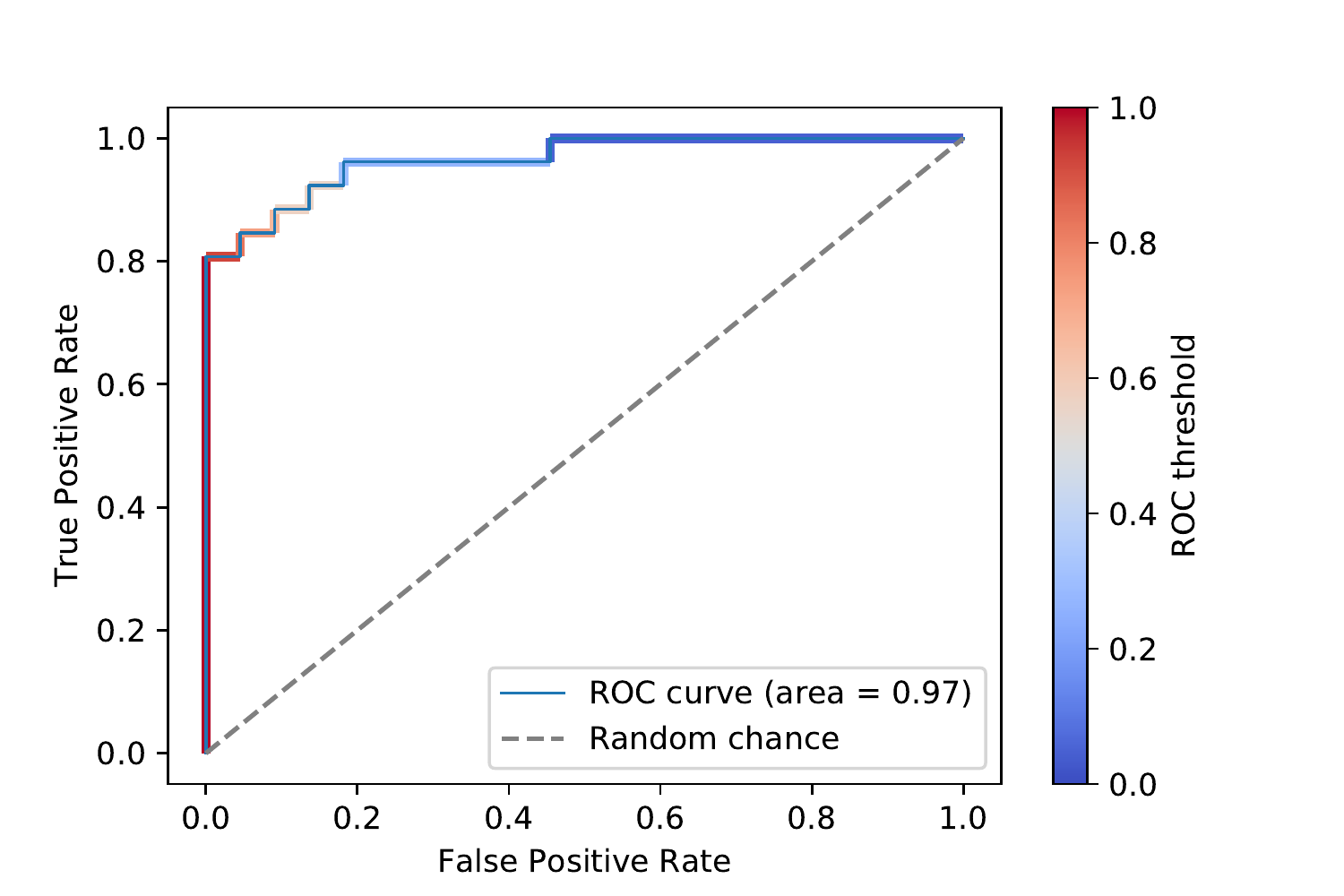}  
\caption{Receiver operating characteristic curve of the cough classifier}
\label{fig:roc}
\end{figure}

The ROC curve of the cough classifier is displayed in Figure \ref{fig:roc}, which users of the COUGHVID database can consult to set a cough detection threshold that suits their specifications. As this figure shows, only 10.4\% of recordings with a \texttt{cough\_detected} value less than 0.8 actually contain cough sounds. Therefore, they should be used only for robustness assessment, and not as valid cough examples. 

\iffalse
Classification results: 89.22\% CV precision on training set, 89.34\% testing balanced accuracy, 96.5\% ROC, 90.56\% F-1 score, 88.89\% testing precision
\fi

\subsection*{Expert Annotation}

To enhance the quality of the dataset with clinically validated information, we were assisted by three expert pulmonologists. Each of them revised 1000 recordings, selecting one of the predefined options to each of the following 10 items:

\begin{enumerate}
 \item \textbf{Quality}: Good; Ok; Poor; No cough present.
 \item \textbf{Type of cough}: Wet (productive); Dry; Can't tell.
 \item \textbf{Audible dyspnea}: Checkbox.
 \item \textbf{Audible wheezing}: Checkbox.
 \item \textbf{Audible stridor}: Checkbox.
 \item \textbf{Audible choking}: Checkbox.
 \item \textbf{Audible nasal congestion}: Checkbox.
 \item \textbf{Nothing specific}: Checkbox.
 \item \textbf{Impression: I think this patient has...}: An upper respiratory tract infection; A lower respiratory tract infection; Obstructive lung disease (Asthma, COPD, ...); COVID-19; Nothing (healthy cough).
 \item \textbf{Impression: the cough is probably...}: Pseudocough / Healthy cough (from a healthy person); Mild (from a sick person); Severe (from a sick person); Can't tell.
\end{enumerate}

As a labeling support tool, we used an online spreadsheet using Google Sheets\textsuperscript{\textcopyright}. Thus, the experts could play the recordings directly inside the browser and select their answers in a convenient way. Once a recording had been labeled, the background color of the full row turned to green for easier navigation. The time required for each expert for labeling the 1000 recordings was around 10 hours, without significant differences among the three experts.

In addition to the personal spreadsheets, the experts were provided with the following general instructions:

\begin{itemize}
 \item \textit{For binary variables (Columns E-J) the box should be ticked for any reasonable suspicion of the sounds being heard.}
 \item \textit{In Column K ("Impression: I think this patient has..."), you should check what you consider the most likely diagnosis, knowing that the records were collected between 01/04/2020 and 18/05/2020.}
 \item \textit{Each row is automatically marked as "completed" after providing an answer to column K. However, please try to give an answer to every column.}
\end{itemize}

Also, the following criteria for assessing quality was indicated to the experts:

\begin{itemize}
 \item \textbf{Good}: \textit{Cough present with minimal background noise.}
 \item \textbf{Ok}: \textit{Cough present with background noise.}
 \item \textbf{Poor}: \textit{Cough present with significant background noise.}
\end{itemize}

The selection of the recordings to be labeled was done through stratified random sampling and after pre-filtering using the automatic cough classifier described above, requiring a minimum probability of 0.8 of containing cough sounds. The stratification was based on the self-reported \texttt{status} variable, as follows:
\begin{itemize}
 \item 25\% of the recordings with \texttt{COVID} value.
 \item 35\% of the recordings with \texttt{symptomatic} value.
 \item 25\% of the recordings with \texttt{healthy} value.
 \item 15\% of the recordings with no reported \texttt{status}.
\end{itemize}

Finally, we ensured that 15\% of the recordings were labeled by all three reviewers, so that we could assess the level of agreement among them.

\section*{Data Records}

% The Data Records section should be used to explain each data record associated with this work, including the repository where this information is stored, and to provide an overview of the data files and their formats. Each external data record should be cited numerically in the text of this section, for example \cite{Hao:gidmaps:2014}, and included in the main reference list as described below. A data citation should also be placed in the subsection of the Methods containing the data-collection or analytical procedure(s) used to derive the corresponding record. Providing a direct link to the dataset may also be helpful to readers (\hyperlink{https://doi.org/10.6084/m9.figshare.853801}{https://doi.org/10.6084/m9.figshare.853801}).
% 
% Tables should be used to support the data records, and should clearly indicate the samples and subjects (study inputs), their provenance, and the experimental manipulations performed on each (please see 'Tables' below). They should also specify the data output resulting from each data-collection or analytical step, should these form part of the archived record.

Each cough recording consists of two files with the same name but different extensions. One of the files contains the audio data, as it was directly received at the COUGHVID servers, and it can be in the WEBM~\cite{Bankoski2011} or OGG~\cite{rfc3533} formats, respectively with the \texttt{.webm} and \texttt{.ogg} extensions. In all cases, the audio codec is Opus~\cite{rfc6716} and the sampling frequency 48 kHz. The second file contains the metadata encoded as plain text in the JSON format~\cite{rfc7159}, and has the \texttt{.json} extension. The file name is a random string generated according to the UUID V4 protocol~\cite{rfc4122}.

In the metadata we may distinguish three types of variables, related to: 1) context information (timestamp and the probability that the recording actually contains cough sounds), 2) self-reported information provided by the user, and 3) the labels provided by expert medical annotators about the clinical assessment of the cough recording. The only processing performed on the metadata was the reduction in the precision of the geolocation coordinates to just one decimal digit to ensure privacy protection. A full description of all the metadata variables is provided in Tables~\ref{tab:metadata} and~\ref{tab:expert_labels}.

As an illustrative example, let us consider the recording with UUID \texttt{4e47612c-6c09-4580-a9b6-2eb6bf2ab40c}. We can see the audio properties using a tool such as \texttt{ffprobe} (included in the \texttt{ffmpeg} software package~\cite{ffmpeg}), while the metadata can be directly displayed as a text file:

\subsection*{4e47612c-6c09-4580-a9b6-2eb6bf2ab40c.webm}
\begin{verbatim}
Input #0, matroska,webm, from '4e47612c-6c09-4580-a9b6-2eb6bf2ab40c.webm':
  Metadata:
    encoder         : Chrome
  Duration: N/A, start: 0.000000, bitrate: N/A
    Stream #0:0(eng): Audio: opus, 48000 Hz, mono, fltp (default)
\end{verbatim}

\subsection*{4e47612c-6c09-4580-a9b6-2eb6bf2ab40c.json}
\begin{verbatim}
{
    "datetime": "2020-04-10T10:30:31.576207+00:00",
    "cough_detected": "0.9466",
    "age": "50",
    "gender": "male",
    "respiratory_condition": "True",
    "fever_muscle_pain": "False",
    "status": "COVID-19",
    "expert_labels_1": {
        "quality": "ok",
        "cough_type": "dry",
        "dyspnea": "False",
        "wheezing": "False",
        "stridor": "False",
        "choking": "False",
        "congestion": "False",
        "nothing": "True",
        "diagnosis": "COVID-19",
        "severity": "mild"
    }
} 
\end{verbatim}

\subsubsection*{}
For convenience, a file compiling all of the available metadata is also provided. This file is named \texttt{metadata\_compiled.csv}, and is a CSV file with 40 columns and one row per record. The first column corresponds to the UUID of each recording, and may be used as an index, while the rest of the columns correspond to the variables described in Tables~\ref{tab:metadata} and~\ref{tab:expert_labels}. The expert annotation variables have been expanded, and are named \texttt{quality\_1, cough\_type\_1 \ldots~diagnosis\_3, severity\_3}.

\section*{Technical Validation}

\subsection*{Demographic representativeness}
An important requirement for large datasets is that they must represent a wide range of subject demographics. Demographic statistics were collected across all recordings that provided metadata and that were classified as coughs with a probability above 0.8 by the XGB cough detection model. There were slightly more male subjects than female (65.5\% and 33.8\%, respectively), and the majority of subjects did not have pre-existing respiratory conditions (81.9\%). The percentages of healthy, COVID-19 symptomatic, and COVID-19 positive subjects were 77\%, 15.5\%, and 7.5\%, respectively. The average age of the recordings was 34.4 years with a standard deviation of 12.8 years. This shows that a wide variety of ages, genders, and health statuses are captured within our dataset.

\subsection*{Geographic representativeness}
In order to assess the plausibility that samples labeled as COVID-19 truly originated from people who tested positive for the disease, we analyzed the geographic locations of the samples for which this data was provided. We then evaluated the COVID-19 statistics of the countries at the time each recording was sent to determine if these countries had high infection rates at the time of recording. Of the 10,743 samples that the XGB model classified as cough sounds, 6,485 of them provided GPS information, 676 of which reported COVID-19 symptoms and 372 claim to have been diagnosed with COVID-19. 

Studies have shown that coughing persists in a significant proportion of COVID-19 cases 14-21 days after receiving a positive PCR test \cite{Tenforde2020}. Therefore, we combined the World Health Organization's statistics on daily new COVID-19 cases \cite{covidStats} with the United Nations 2019 population database \cite{populationData} to determine the rate of new infections in the country from which each recording originated in the 14 days prior to it being uploaded to our web server. This analysis revealed that 94.4\% of our recordings labeled as COVID-19 came from countries with more than 20 newly-confirmed COVID-19 cases per 1 million people. Similarly, 91.3\% of recordings labeled as symptomatic originated from these countries.

\begin{figure}[ht]
  \centering
  % include first image
  \includegraphics[width=\linewidth]{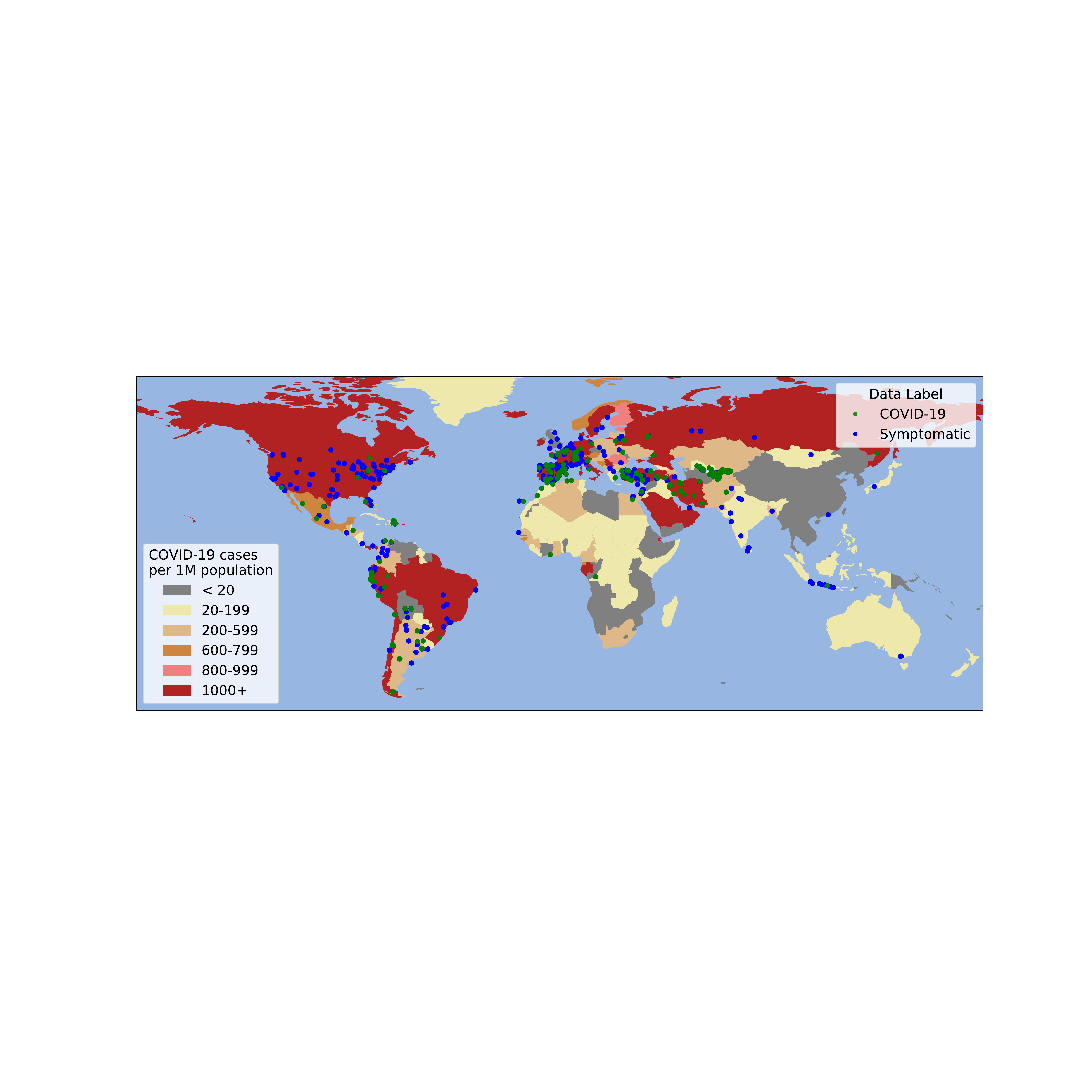}  
\caption{Cumulative COVID-19 cases in April and May 2020 per 1 million population, along with the GPS coordinates of the received recordings}
\label{fig:covid_map}
\end{figure}

Figure \ref{fig:covid_map} shows a map of the world with countries color coded according to the cumulative COVID-19 positive tests in April and May 2020 per 1 million population. We also show the COVID-19 and symptomatic recordings providing GPS information that were collected within this time period. This figure shows that most recordings were sent from countries with moderate-to-high COVID-19 infection rates at the time.

\subsection*{Inter-Rater Reliability}
In order to determine the extent to which the three expert pulmonologists agreed on their cough sound labeling, Fleiss' Kappa scores\cite{Fleiss1971}, $K_{Fleiss}$, were computed for each question among the 150 common recordings. The results are displayed in Table \ref{tab:agreement}. This analysis revealed moderate agreement on audible nasal congestion, fair agreement on the type of cough, as well as slight agreement on the cough severity, nothing specific, audible wheezing, and audible dyspnea. 

\begin{figure}[ht]
  \centering
  % include first image
  \includegraphics[width=\linewidth]{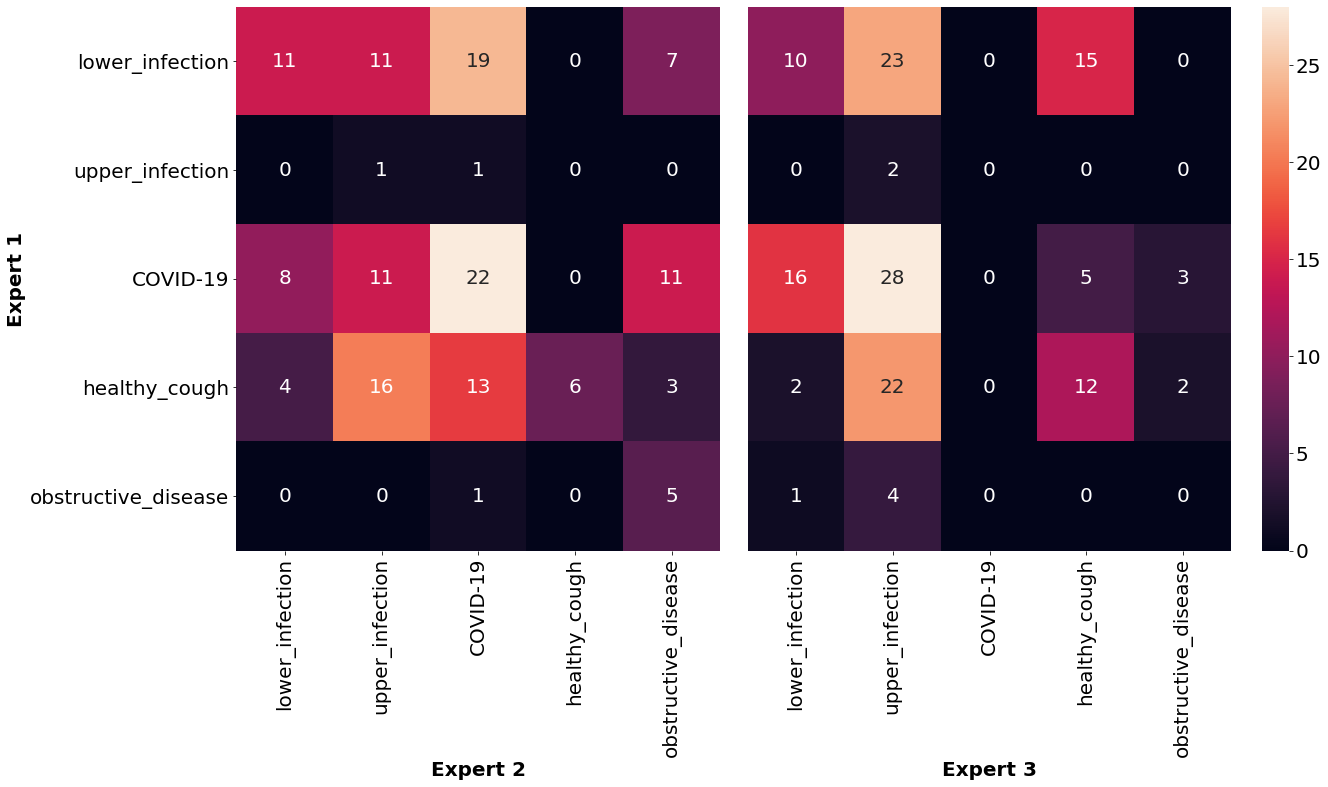}  
\caption{Confusion matrix of common cough recording diagnoses between pulmonology experts.}
\label{fig:conf_mat}
\end{figure}

There was a poor agreement among experts on the cough diagnosis ($K_{Fleiss} = 0.0031$), which is reflective of the fact that COVID-19 symptomology includes symptoms of both upper respiratory tract infections (e.g., rhinorrhea, sore throat, etc.) and lower respiratory tract infections (e.g., pneumonia, ground-glass opacities, etc.) \cite{Rothan2020}. There was a slight agreement between Experts 1 and 2 on the diagnosis ($K_{Fleiss} = 0.0774$), whereas Expert 3 did not label any of the common recordings as COVID-19. Out of the 86 coughs that at least one rater labeled as COVID-19, 22 had a majority consensus. When the majority agreed that a cough was COVID-19, Expert 3 diagnosed it as an upper respiratory infection 59.1\% of the time, a lower respiratory tract infection 27.3\% of the time, and nothing 13.6\% of the time. A confusion matrix between Experts 1 and 3 is displayed in Figure \ref{fig:conf_mat}, which shows a higher degree of agreement between Experts 1 and 2 than either with Expert 3. 

\subsection*{Trends in Expert COVID-19 Cough Labeling}

All of the coughs labeled as COVID-19 among the three experts were subsequently pooled together and analyzed for trends in the attributes of the cough recordings. In the case of overlaps, the diagnoses of Expert 1 were taken into account, producing 632 total COVID-19-labeled cough records. The vast majority of coughs do not exhibit audible dyspnea (93.0\%), wheezing (90.5\%), stridor (98.7\%), choking (99.1\%), or nasal congestion (99.2\%). Additionally, 87.3\% of COVID-19-labeled coughs are annotated as dry, which is consistent with literature stating that a dry cough is a common COVID-19 symptom \cite{Rothan2020, whochina}. Finally, 86.2\% of these coughs are labeled as mild. These commonalities among COVID-19 labeled coughs reflect the consistency of the database.

\iffalse
This section presents any experiments or analyses that are needed
to support the technical quality of the dataset. This section may
be supported by figures and tables, as needed. This is a required
section; authors must present information justifying the reliability
of their data.
\fi

\section*{Private Set and Testing Protocol}

% The Usage Notes should contain brief instructions to assist other researchers with reuse of the data. This may include discussion of software packages that are suitable for analysing the assay data files, suggested downstream processing steps (e.g. normalization, etc.), or tips for integrating or comparing the data records with other datasets. Authors are encouraged to provide code, programs or data-processing workflows if they may help others understand or use the data. Please see our code availability policy for advice on supplying custom code alongside Data Descriptor manuscripts.
% 
% For studies involving privacy or safety controls on public access to the data, this section should describe in detail these controls, including how authors can apply to access the data, what criteria will be used to determine who may access the data, and any limitations on data use. 

In order to ensure the reproducibility of the experimental results that use the COUGHVID crowdsourcing dataset, a private test set has been kept out from publishing. Recordings in this private set have been randomly selected from those having at least labels from one expert.

The evaluation of models on the private test set will be open to the entire scientific community, but to ensure a fair use, the performance measurements will be obtained by an independent evaluator. Any researcher that demonstrates promising results on the public dataset using cross-validation may apply for an independent evaluation on the private test set according to the protocol described in the data repository. This protocol shall be regularly updated in line with the available technology to ensure that it is as convenient as possible for all parties.

Since one of the aims of this project is to go beyond the study of COVID-19, every variable except \texttt{datetime} and \texttt{cough\_detected} may be considered the target of a prediction model. This opens the possibility to study several different problems within the same dataset, from just cough identification and sound quality assessment to the detection of different conditions, or even the estimation of age or gender. Conversely, a prediction model may require as input not just the sound recording, but also other metadata variables to provide the necessary context, such as the age or gender of the subject. The restrictions on the sets of variables that can be used as input or as a result of the algorithms will be kept to a minimum.

\section*{Code availability}

The aforementioned XGB classifier used to remove non-cough recordings and feature extraction source code are available on our public \href{https://c4science.ch/diffusion/10770/}{c4science repository}.

\iffalse
For all studies using custom code in the generation or processing of datasets, a statement must be included under the heading "Code availability", indicating whether and how the code can be accessed, including any restrictions to access. This section should also include information on the versions of any software used, if relevant, and any specific variables or parameters used to generate, test, or process the current dataset. 
\fi

\bibliography{sample}

\iffalse
\noindent LaTeX formats citations and references automatically using the bibliography records in your .bib file, which you can edit via the project menu. Use the cite command for an inline citation, e.g. \cite{Kaufman2020, Figueredo:2009dg, Babichev2002, behringer2014manipulating}. For data citations of datasets uploaded to e.g. \emph{figshare}, please use the \verb|howpublished| option in the bib entry to specify the platform and the link, as in the \verb|Hao:gidmaps:2014| example in the sample bibliography file. For journal articles, DOIs should be included for works in press that do not yet have volume or page numbers. For other journal articles, DOIs should be included uniformly for all articles or not at all. We recommend that you encode all DOIs in your bibtex database as full URLs, e.g. https://doi.org/10.1007/s12110-009-9068-2 and then include the following command to remove the default DOI prefix \verb|\newcommand{\doiprefix}{}|.
\fi 

\section*{Acknowledgements} 

This work has been supported in part by the DeepHealth Project (GA No. 825111), and in part by the Swiss NSF ML-Edge Project (GA No. 182009). We would also like to acknowledge Dr. Constantin Bondolfi, Dr. Pierre-Yves Ryser and Dr. Erin Gonvers for their hard work in the manual annotation of the recordings, and also Dr. Mary-Anne Hartley for her invaluable help in the definition of the annotation protocol.

\iffalse
(not compulsory)
Acknowledgements should be brief, and should not include thanks to anonymous referees and editors, or effusive comments. Grant or contribution numbers may be acknowledged.
\fi

\section*{Author contributions statement}

L.O. and T.T. devised the idea and research. L.O. developed the cough signal processing code, created the cough detection model, and performed data analysis on the user metadata and expert labels. T.T. deployed and managed the data collection website, coordinated expert labeling, and compiled all of the metadata. All authors contributed to the writing and editing of the manuscript.

\iffalse
Must include all authors, identified by initials, for example:
A.A. conceived the experiment(s),  A.A. and B.A. conducted the experiment(s), C.A. and D.A. analysed the results.  All authors reviewed the manuscript. 
\fi

\section*{Competing interests} %(mandatory statement)

%The corresponding author is responsible for providing a \href{https://www.nature.com/sdata/policies/editorial-and-publishing-policies#competing}{competing interests statement} on behalf of all authors of the paper. This statement must be included in the submitted article file.

The authors declare no competing interests

\section*{Figures \& Tables}

\begin{table}
\renewcommand{\arraystretch}{1.2}
\begin{center}
\begin{tabularx}{\textwidth}{|l|l|X|X|}
\hline
\textbf{Name} & \textbf{Mandatory} & \textbf{Range of possible values} & \textbf{Description} \\
\hline
\texttt{datetime} & Yes & UTC date and time in ISO 8601 format &  Timestamp of the received recording.\\
\hline
\texttt{cough\_detected} & Yes & Floating point in the interval $[0, 1]$ & Probability that the recording contains cough sounds, according to the automatic detection algorithm described in the Methods section. \\
\hline
\texttt{latitude} & No & Floating point value & Self-reported latitude geolocation coordinate with reduced precision.\\
\hline
\texttt{longitude} & No & Floating point value & Self-reported longitude geolocation coordinate with reduced precision.\\
\hline
\texttt{age} & No & Integer value & Self-reported age value.\\
\hline
\texttt{gender} & No & \texttt{\{female, male, other\}} & Self-reported gender.\\
\hline
\texttt{respiratory\_condition} & No & \texttt{\{True, False\}} & The patient has other respiratory conditions (self-reported).\\
\hline
\texttt{fever\_muscle\_pain} & No & \texttt{\{True, False\}} & The patient has fever or muscle pain (self-reported).\\
\hline
\texttt{status} & No & \texttt{\{COVID, symptomatic, healthy\}} & The patient self-reports that has been diagnosed with COVID-19 (\texttt{COVID}), that has symptoms but no diagnosis (\texttt{symptomatic}), or that is healthy (\texttt{healthy}).\\
\hline
\texttt{expert\_labels\_\{1,2,3\}} & No & JSON dictionary with the manual labels from expert 1, 2 or 3 & The expert annotation variables are described in Table~\ref{tab:expert_labels}.\\
\hline
\end{tabularx}
\end{center}
\caption{\label{tab:metadata}Metadata variables, as they appear in the JSON files.}
\end{table}    

\begin{table}
\renewcommand{\arraystretch}{1.2}
\begin{center}
\begin{tabularx}{\textwidth}{|l|X|X|}
\hline
%Quality_1,Cough_type_1,Dyspnea_1,Wheezing_1,Stridor_1,Choking_1,Congestion_1,Nothing_1,Diagnosis_1,Severity_1
\textbf{Name} & \textbf{Range of possible values} & \textbf{Description} \\
\hline
\texttt{quality} & \texttt{\{good, ok, poor, no\_cough\}} & Quality of the recorded cough sound.\\
\hline
\texttt{cough\_type} & \texttt{\{wet, dry, unknown\}} & Type of cough.\\
\hline
\texttt{dyspnea} & \texttt{\{True, False\}} & Audible dyspnea.\\
\hline
\texttt{wheezing} & \texttt{\{True, False\}} & Audible wheezing.\\
\hline
\texttt{stridor} & \texttt{\{True, False\}} & Audible stridor.\\
\hline
\texttt{choking} & \texttt{\{True, False\}} & Audible choking.\\
\hline
\texttt{congestion} & \texttt{\{True, False\}} & Audible nasal congestion.\\
\hline
\texttt{nothing} & \texttt{\{True, False\}} & Nothing specific is audible.\\
\hline
\texttt{diagnosis} & \texttt{\{upper\_infection, lower\_infection, obstructive\_disease, COVID-19, healthy\_cough\}} & Impression of the expert about the condition of the patient. It can be an upper or lower respiratory tract infection, an obstructive disease (Asthma, COPD, etc), COVID-19, or a healthy cough.\\
\hline
\texttt{severity} & \texttt{\{pseudocough, mild, severe, unknown\}} & Impression of the expert about the severity of the cough. It can be a pseudocough from a healthy patient, a mild or severe cough from a sick patient, or unknown if the expert can't tell.\\
\hline
\end{tabularx}
\end{center}
\caption{\label{tab:expert_labels}Variables provided by the expert annotators.}
\end{table}

\begin{table}[]
\centering
\caption{\label{tab:agreement}Inter-Expert Label Consistency}
\begin{tabular}{|l|l|l|}
\hline
Item        & $K_{Fleiss}$ & Agreement \\ \hline
quality     & -0.12   & Poor      \\ \hline
cough\_type & 0.23    & Fair      \\ \hline
dyspnea     & 0.04    & Slight    \\ \hline
wheezing    & 0.04    & Slight    \\ \hline
stridor     & -0.01   & Poor      \\ \hline
choking     & -0.01   & Poor      \\ \hline
congestion  & 0.49    & Moderate  \\ \hline
nothing     & 0.10    & Slight    \\ \hline
diagnosis   & 0.00    & Poor      \\ \hline
severity    & 0.12    & Slight    \\ \hline
\end{tabular}
\end{table}

\end{document}